\documentclass[fleqn,twoside]{article}
\usepackage[headings]{espcrc2}

\readRCS
$Id: espcrc2.tex,v 1.2 2004/02/24 11:22:11 spepping Exp $
\usepackage{graphicx}
\usepackage[figuresright]{rotating}

\newcommand{\AmS}{{\protect\the\textfont2
  A\kern-.1667em\lower.5ex\hbox{M}\kern-.125emS}}

\title{Spectroscopy of Heavy Quarkonia
 \thanks{Presented at the 6th International Conference on Hyperons, Charm
           and Beauty Hadrons, Illinois Institute of Technology, Chicago, USA}}

\author{Holger St\"ock
        \address{University of Florida, Department of Physics, PO Box 118440,
                 Gainesville, FL 32611-8440, USA}}
       
\runtitle{Spectroscopy of Heavy Quarkonia}
\runauthor{H. St\"ock}

\begin{document}

\begin{abstract}
In the last few years, the CLEO III experiment has recorded a large collection 
of data sets at the $\Upsilon(1S)$, $\Upsilon(2S)$, $\Upsilon(3S)$ and 
$\psi(2S)$ resonances. Preliminary results of studies of these data sets are
shown here, which include the observation of a new $\Upsilon(1D)$ state, as
well as several hadronic and radiative transitions of $\Upsilon$ and $\psi(2S)$
states. In addition, precision branching ratio measurements of charmonium and
bottomonium states and recent developments involving the $\eta_c(2S)$ and
$X(3872)$ states are presented.

\vspace{1pc}
\end{abstract}

\maketitle

\section{INTRODUCTION}

Heavy quarkonia - $c\bar{c}$ and $b\bar{b}$ bound states - are still a very
rich exploration ground. For example, no $b\bar{b}$ singlet states have yet
been found, and only a few hadronic and radiative decays are known.

The masses of charm and bottom quarks are relatively large ($\approx$1.5 and 
$\approx$4.5 GeV). As a consequence, the velocities of these quarks in hadrons
are non-relativistic and the strong coupling constant $\alpha_s$ is quite small
($\approx$0.3 for $c\bar{c}$ and $\approx$0.2 for $b\bar{b}$). Hence, heavy
quarkonia provide the best means of testing the theories of strong interaction,
i.e. QCD in both pertubative and non-pertubative regimes, QCD inspired purely
phenomenological potential models, NRQCD and lattice QCD.

In this paper, the latest results on charmonium and bottomonium spectroscopy
from the CLEO III experiment are presented and recent developments of the
searches for the $\eta_c(2S)$ and $X(3872)$ are discussed.

\section{THE CLEO III EXPERIMENT AND DATA SETS}

The data used in the presented studies were taken with the CLEO III detector at
the CESR $e^+e^-$ storage ring. The detector includes a silicon microvertex
detector, a drift chamber and a ring imaging cerenkov detector (RICH), as well
as a crystal calorimeter. On the outside, the detector is surrounded by muon
chambers. The tracking volume is placed in a uniform 1.5 T solenoidal magnetic 
field.

During the last two years, the CLEO III experiment recorded data at the 
$\Upsilon (1S)$, $\Upsilon (2S)$, $\Upsilon (3S)$ and $\psi(2S)$ resonances,
resulting in approximately 38 million hadronic events. In addition, data were
also recorded below the peak of each resonance for background purposes and for
scans across the resonances. Table \ref{table:datasets} summarizes the data
sets used.

\begin{table*}[htb]
\caption{Summary of the CLEO III data sets used in the studies which are
         presented in this paper. For comparison, in brackets the number of
         events recorded with the CLEO II detector are shown.}
\label{table:datasets}
\renewcommand{\tabcolsep}{2.35pc} 
\renewcommand{\arraystretch}{1.2} 
\begin{tabular}{@{}c|c|c|c}
\hline
$E_{cm}$ (GeV) & Resonance      & \# Events ($10^6$) & Experiment \\
\hline
 9.46          & $\Upsilon(1S)$ & 20 (2)             & CLEO III (CLEO II) \\
10.02          & $\Upsilon(2S)$ & 10 (0.5)           & CLEO III (CLEO II) \\
10.36          & $\Upsilon(3S)$ &  5 (0.5)           & CLEO III (CLEO II) \\
 3.69          & $\psi(2S)$     &  3                 & CLEO III \\
\hline
\end{tabular}
\end{table*}

\section{BOTTOMONIUM SPECTROSCOPY}

\subsection{First Observation of $\Upsilon(1^3D_2)$ State}

CLEO has made the first observation of the bottomonium state $\Upsilon(1^3D_2)$
\cite{bonvicini0}. This state was produced in a two-photon cascade via the
$\chi(2P)$ state starting from the $\Upsilon(3S)$ resonance:
$\Upsilon(3S) \rightarrow \gamma \chi(2P) 
              \rightarrow \gamma \gamma \Upsilon(1^3D_2)$

To suppress photon backgrounds from $\pi^0$s, which are copiously produced in
gluonic annihilation of the $b\bar{b}$ states, events with two or more
subsequent photon transitions were selected via the cascade
$\Upsilon(1^3D_2) \rightarrow \gamma \chi(1P)
                  \rightarrow \gamma \gamma \Upsilon(1S)$
followed by the $\Upsilon(1S)$ annihilation into either $e^+e^-$ or 
$\mu^+\mu^-$.

In this four-photon cascade $34.5 \pm 6.4$ signal events were observed which
translate  into a significance of $10.2 \sigma$. The $\Upsilon(1^3D_2)$ mass
was measured to $M = 10161.1 \pm 0.6(stat) \pm 1.6(syst) ~MeV$ and the product
branching ratio was determined to 
${\cal{B}}(4\gamma\ell^+\ell^-) = 
          2.6 \pm 0.5(stat) \pm 0.5(syst) \times 10^{-5}$

The measured mass is in good agreement with the mass of the $\Upsilon(1^3D_2)$
state predicted by lattice QCD calculations \cite{lepage0} and those potential
models which give a good fit to the other known $b\bar{b}$ states 
\cite{skwarnicki0}. Also, a prediction of the product branching ratio of
$3.76 \times 10^{-5}$ by Godfrey and Rosner \cite{godfrey0} is in good
agreement with our branching ratio measurement.

\subsection{Measurement of the Muonic Branching Ratio of 
            ${\cal{B}}(\Upsilon(nS)$ Resonances}

Previous measurements have established ${\cal{B}}_{\mu\mu}$ with a 2.4\%
accuracy for the $\Upsilon(1S)$ \cite{PDG}, and a modest 16\% accuracy for the 
$\Upsilon(2S)$ \cite{haas0,albrecht0,kaarsberg0,kobel0} and $\Upsilon(3S)$
\cite{kaarsberg0,andrews0,chen0}. CLEO has made new measurements of 
${\cal{B}}_{\mu\mu}$ in all three resonances using a much larger data set
together with a more advanced detector.

To determine ${\cal{B}}_{\mu\mu}$, we measured 
$\bar{\cal{B}} \equiv \Gamma_{\mu\mu} / \Gamma_{had}$, where 
$\Gamma_{\mu\mu}$ ($\Gamma_{had}$) is the rate for $\Upsilon$ decay to
$\mu^+\mu^-$ (hadron). $\Gamma_{had}$ includes all decay modes other than the
electromagnetic decays to $e^+e^-$, $\mu^+\mu^-$, and $\tau^+\tau^-$.
Assuming lepton universality, we have
${\cal{B}}_{\mu\mu} = \Gamma_{\mu\mu} / \Gamma = 
      \bar{\cal{B}}_{\mu\mu} / (1+3\bar{\cal{B}}_{\mu\mu})$.

The results of the measurements are shown in Table \ref{table:YnStomumu}. The
measurement for $\Upsilon(1S)$ is in good agreement with the world average. The
obtained branching ratios for the $\Upsilon(2S)$ and $\Upsilon(3S)$
resonances are significantly larger than prior measurements and the world 
average values, resulting in a narrower total decay widths.

\begin{table*}[htb]
\caption{Summary of the branching ratio measurements of
         ${\cal{B}}(\Upsilon(nS) \rightarrow \mu^+\mu^-)$. For comparison, the
         world averages (PDG) \cite{PDG} are given.}
\label{table:YnStomumu}
\renewcommand{\tabcolsep}{2pc} 
\renewcommand{\arraystretch}{1.2} 
\begin{tabular}{@{}c|c|c|c}
\hline
${\cal{B}}_{\mu\mu}$ (\%) & $\Upsilon(1S)$ & $\Upsilon(2S)$ & $\Upsilon(3S)$ \\
\hline
CLEO                 & 2.53 $\pm$ 0.02 $\pm$ 0.05
                     & 2.11 $\pm$ 0.03 $\pm$ 0.05
                     & 2.44 $\pm$ 0.07 $\pm$ 0.05 \\
PDG                  & 2.48 $\pm$ 0.06 & 1.31 $\pm$ 0.21 & 1.81 $\pm$ 0.17 \\
\hline
\end{tabular}
\end{table*}

\subsection{J/$\psi$ Production in $\Upsilon(1S)$ Decays}

In $\Upsilon$ decays charm should be produced through the color octet mechanism
\cite{braaten0}. To study this mechanism, CLEO searched for the inclusive
production of the J/$\psi$ in $\Upsilon(1S)$ resonance decays with $J/\psi$
decaying into $e^+e^-$ and $\mu^+\mu^-$. Additional cuts were used to suppress
radiative returns to the J/$\psi$ and $\psi(2S)$. As a cross check,
the same analysis method was used on $\Upsilon(4S)$ data to verify that its
results were as expected. The branching ratios for the two lepton modes are
averaged, thereby obtaining a preliminary branching ratio of
${\cal{B}}(\Upsilon(1S) \rightarrow J/\psi + ~X) =
     6.4 \pm 0.4(stat) \pm 0.6(syst) \times 10^{-4}$.

Also of interest is the beam energy scaled momentum spectrum of the J/$\psi$'s 
in this process. It appears that this spectrum is softer than what is expected
from a naive color octet model, although it might be possible to address this
issue with the emission of soft gluons in the theoretical calculations.

\subsection{Di-Pion Transitions from $\Upsilon(3S)$}

CLEO also performed studies of charged and neutral di-pion transitions from the
$\Upsilon(3S)$ resonance to $\Upsilon(2S)$ and $\Upsilon(1S)$. For the neutral
di-pion transitions of $\Upsilon(3S)$ to $\Upsilon(2S)$ and $\Upsilon(1S)$ we
measure the preliminary branching ratios as \\
\noindent
${\cal{B}} (\Upsilon(3S) \rightarrow \pi^0\pi^0~\Upsilon(2S)) = \\
~~~~~~~~~~~~~~~~~~~~~~~~~2.02 \pm 0.18(stat) \pm 0.38(syst) \%$ \\
${\cal{B}} (\Upsilon(3S) \rightarrow \pi^0\pi^0~\Upsilon(1S)) = \\
~~~~~~~~~~~~~~~~~~~~~~~~~1.88 \pm 0.08(stat) \pm 0.31(syst) \%$

The $\pi^0\pi^0$ effective mass spectrum for the transition to the
$\Upsilon(2S)$ has a shape consistent with several theoretical predictions.
The mass spectrum for the transition to the $\Upsilon(1S)$ has a ``double
humped'' shape which was also observed in charged di-pion transitions
\cite{butler0}. Further measurements of neutral and the charged di-pion
transitions are in progress.

\subsection{Radiative Transitions from $\Upsilon(nS)$}

Radiative transitions from $\Upsilon(nS)$ states are an 
\newpage
\noindent
excellent place to look for new intermediate resonances, such as $b\bar{b}$
singlet states.

In M1 transitions, CLEO has searched for the states $\eta_b(1S)$ and 
$\eta_b(2S)$. No significant signals were found and only upper limits for
branching ratios as a function of the photon energy were determined.

In E1 transitions, we observed the following three transitions: \\
$\Upsilon(3^3S_1) \rightarrow \gamma ~\chi_b(1^3P_J)$ \\
$\chi_b(1^3P_0) \rightarrow \gamma ~ \Upsilon(1^3S_1)$ \\
$\chi_b(2^3P_0) \rightarrow \gamma ~ \Upsilon(2^3S_1)$

No preliminary branching ratios are available at this time. Further precision
measurements are in progress.

\section{CHARMONIUM SPECTROSCOPY}

\subsection{Radiative Transitions from $\psi(2S)$}

CLEO has measured the branching ratios for the radiative transitions
$\psi(2S)\rightarrow \gamma\chi_c(1P_J)$ and 
$\psi(2S) \rightarrow \gamma\eta_c(1S)$. Table \ref{table:Psi2SradTrans}
summarizes the four measurements. The ratios are in good agreement with the
Particle Data Group (PDG) averages. CLEO also confirms the M1 transition to
$\eta_c(1S)$ made by Crystal Ball. However, we find no indication of the M1
transition to $\eta_c(2S)$ which was reported by the Crystal Ball collaboration
20 years ago.
\begin{table*}[htb]
\caption{Summary of the branching ratio measurements for
         $\psi(2S)$ radiative transitions. For comparison,
         the PDG values \cite{PDG} are given.}
\label{table:Psi2SradTrans}
\renewcommand{\tabcolsep}{0.75pc} 
\renewcommand{\arraystretch}{1.2} 
\begin{tabular}{@{}c|c|c|c|c}
\hline
${\cal{B}}$ (\%) & \multicolumn{3}{c|}{$\psi(2S)\rightarrow\gamma\chi_c(1P_J)$}
                 & $\psi(2S) \rightarrow \gamma\eta_c(1S)$ \\
                 & \multicolumn{1}{c}{J = 2 (E1 line)} 
                 & \multicolumn{1}{c}{J = 1 (E1 line)}
                 & \multicolumn{1}{c|}{J = 0 (E1 line)}
                 & \multicolumn{1}{c}{J = 0 (M1 line)} \\
\hline
CLEO      & 9.75 $\pm$ 0.14 $\pm$ 1.17
          & 9.64 $\pm$ 0.11 $\pm$ 0.69
          & 9.83 $\pm$ 0.13 $\pm$ 0.87 
          & 0.278 $\pm$ 0.033 $\pm$ 0.049 \\
PDG       & 7.8 $\pm$ 0.8 & 8.7 $\pm$ 0.8 & 9.3 $\pm$ 0.8 & 0.28 $\pm$ 0.06 \\
\hline
\end{tabular}
\end{table*}

\subsection{Experimental Results on $\eta_c(2S)$}
The $\eta_c'$ has a very colorful history with one observation followed by a
number of fruitless searches. This situation changed recently with the Belle
observation of robust $\eta_c'$ signals in the two channels
$B^\pm \rightarrow K^\pm \eta_c' \rightarrow K^\pm K_s K^\pm \pi^\mp$ with
$M = 3654 \pm 6(stat) \pm 8(syst) ~MeV$ and $e^+e^- \rightarrow J/\psi \eta_c$
with $M = 3622 \pm 12 ~MeV$ \cite{choi0,abe0}, quickly followed by BaBar and
CLEO observations of $\eta_c'$ signals in two-photon fusion processes. BaBar
reported a mass of $M = 3630.8 \pm 3.4(stat) \pm 1.0(syst) ~MeV$ and a width of
$\Gamma = 17.0 \pm 8.3(stat) \pm 2.5(syst) ~MeV$ \cite{aubert0}. CLEO measured
the mass to $M = 3642.9 \pm 3.1(stat) \pm 1.5(syst) ~MeV$ and the width to
$\Gamma < 31 MeV$ (90\% C.L.) \cite{asner0}. The CLEO observation was made in
two data sets with substantially different detectors and software systems. The
results of the three experiments are in reasonable agreement.

Using a combined mass value (Belle, BaBar, CLEO) of $M = 3637 \pm 4.4 ~MeV$,
the value of the hyperfine mass splitting between the $\psi(2S)$ and the
$\eta_c'$ can be given as $\Delta M(2S) = 48.6 \pm 4.4 ~MeV$. This value for
$\Delta M(2S)$ is much smaller than most theoretical predictions and should
lead to a new insight into coupled channel effects and spin-spin contribution
of the confinement part of the $q\bar{q}$ potential.

\subsection{Search for the Narrow State $X(3872)$}

This new narrow state was first observed by the Belle collaboration in the decay
channel $B^\pm \rightarrow K^\pm ~X(3872)$ with
$X(3872) \rightarrow \pi^+\pi^- J/\psi$ and $J/\psi$ decaying into a lepton
pair \cite{choi1}. The reported mass and width are
$M = 3872.0 \pm 0.6(stat) \pm 0.5(syst) ~MeV$ and $\Gamma < 2.3 ~MeV$ (90\%
C.L.). The CDF and D0 collaborations have confirmed the $X(3872)$ observation
in proton-antiproton annihilation
$p\bar{p} \rightarrow X(3872) + X$ with $X(3872) \rightarrow \pi^+\pi^- J/\psi$
and $J/\psi$ decaying into a  muon pair \cite{acosta0,abazov0}. The reported
masses are $M = 3871.3 \pm 0.7(stat) \pm 0.4(syst) ~MeV$ (CDF) and
$M = 3871.8 \pm 3.1(stat) \pm 3.0(syst) ~MeV$ (D0).

CLEO searched for the $X(3872)$ state in untagged two-photon fusion processes
and initial-state-radiation production, analyzing the exclusive channels
$X(3872) \rightarrow \pi^+\pi^- J/\psi$ with $J/\psi$ decaying into a lepton
pair. No signal was found, but upper limits could be set. For untagged
two-photon fusion processes, we find \\
$(2J+1) \Gamma_{\ell\ell} {\cal{B}}(X \rightarrow \pi^+\pi^- J/\psi) < 16.7 ~eV$
(90\% C.L.) and for ISR production we report
$\Gamma_{ee} {\cal{B}}(X \rightarrow \pi^+\pi^- J/\psi) < 6.8 ~eV$ (90\% C.L.).

\section{SUMMARY AND CONCLUSION}

CLEO is now fully exploiting the world's largest sample of $\Upsilon$ decays. 
We have reported on the observation of the $\Upsilon(1^3D_2)$ state and several
hadronic and radiative transitions of $\Upsilon$ and $\psi(2S)$ states. We also
measured the rate for inclusive $\psi$ production in $\Upsilon(1S)$ decays. In
addition, we presented precision branching ratio measurements and searches for
$\eta_c(2S)$ and $X(3872)$. All results are preliminary.

As demonstrated above, spectroscopy of heavy quarkonia is a very active field.
Collections of large data samples are now available from CLEO III ($b\bar{b}$), 
BES II ($c\bar{c}$) and CLEO-c ($c\bar{c}$). Many new important experimental
observations and measurements have emerged and many more are expected for the
near future.

\section*{ACKNOWLEDGMENTS}

We gratefully acknowledge the effort of the CESR staff in providing our
experiment with excellent luminosity and running conditions. This work was
supported by the National Science Foundation and the U.S. Department of Energy.

\end{document}